\documentstyle[12pt]{article}

\font\l=cmr10 at 10pt
\font\ls=cmr7
\font\lss=cmr5
\font\lsy=cmsy10
\font\lsys=cmsy7
\font\lsyss=cmsy5
\font\lmi=cmmi10
\font\lmis=cmmi7
\font\lmiss=cmmi5
\font\lex=cmex10
 at 6truept
 at 6truept
\title{Signature of relic heavy stable neutrinos in underground experiments}

\author{D.Fargion$^{(1)(3)(7)}$, M.Yu.Khlopov$^{(4)(5)(6)}$,
R.V.Konoplich$^{(1)(4)(5)(6)}$\\
and R.Mignani$^{(2)(3)}$\\
\,\,\\
\,\,\\
$^{(1)}$ Dipartimento di Fisica\\
Universit\`a degli Studi ``La Sapienza''\\
P.le A.Moro, 2 - 00185 Roma, Italy\\
$^{(2)}$ Dipartimento di Fisica ``E.Amaldi''\\
Universit\`a degli Studi ``Roma Tre''\\
Via della Vasca Navale, 84 - 00146 Roma, Italy\\
$^{(3)}$ I.N.F.N. Sezione di Roma I\\
c/o Dipartimento di Fisica\\
Universit\`a degli Studi ``La Sapienza''\\
P.le A.Moro, 2 - 00185 Roma, Italy\\
$^{(4)}$ Center for Cosmoparticle Physics - Cosmion''\\
Miusskaya Pl., 4, 125047 Moscow, Russia\\
$^{(5)}$ Institute of Applied Mathematics M.V.Keldysh\\
Miusskaya Pl., 4, 125047 Moscow, Russia\\
$^{(6)}$ Moscow Engineering Physics Institute\\
Kashirskoe sh,m 31, 115409 Moscow, Russia\\
$^{(7)}$ E.E. Faculty Technion Institute, Haifa, Israel}

\begin{document}
\baselineskip=24pt
\maketitle
\vfill\eject
\noindent
{\bf Abstract}
\vskip 6pt

Considering heavy stable neutrinos of 4$^{th}$ generation we calculate the relic
density of such neutrinos in the Universe. Taking into account the condensation of
heavy neutrinos in the Galaxy and applying the results of calculations to
experimental data from underground experiments on search for WIMPs in elastic
neutral current scattering on nuclei we found an exclusion region of neutrino mass
60 GeV $<$ m $<$ 290 GeV. The bounds obtained from present underground experiments while
confirming the previous bounds derived from analysis of cosmic ray spectra are more
reliable ones. We discuss also the first indication of elastic scattering induced by
WIMP in DAMA experiment finding a very narrow window of neutrino mass 45 GeV $<$ m $<$
50 GeV compatible with the possible signal rate in the detector.
\vfill\eject

The are are strong theoretical arguments and experimental evidences favoring the
average density of the matter in the Universe, which might be significantly greater
than allowable 15$\%$ [1] of critical density possibly provided by baryons. The
nature of this dark matter is one of the most important questions facing both
cosmology and particle physics. Amongst the variety of dark matter candidates the
theoretically favorite candidates are light neutrinos, axions and neutralinos.
However in some models heavy neutrinos can play an important role in cosmology as a
cold dark matter contributing a part of closure density.

In the present article we calculate the heavy neutrino density in the Galaxy and
apply the results of calculations to experimental data from underground experiments
on the direct search for WIMP - nucleus elastic scattering in order to obtain bounds
on the mass of neutrino of 4$^{th}$ generation.

In order to be specific we consider the standard electroweak model, including,
however, one additional family of fermions. The heavy neutrino $\nu$ and heavy charged
lepton L form a standard SU(2)$_L$ doublet. In order to ensure the stability of the
heavy neutrino, we assume that its mass m $<$ M$_L$ and that the heavy neutrino is a
Dirac particle.

It is known that modern laboratory experimental results are not inconsistent with the
existence of heavy Dirac neutrinos with mass m $<$ M$_Z$/2, where M$_Z$ is the mass of
the Z boson. In the early Universe at high temperatures such heavy neutrinos should
be in thermal equilibrium with other species of particles. As the temperature in the
Universe drops, heavy neutrinos become nonrelativistic (at T=m) and their abundance
falls off rapidly according to exponential law. In the further expansion of the
Universe, as the temperature decreases below the freeze-out value T$_f$, the weak
interaction processes become too slow to keep neutrinos in equilibrium with other
particles. As a consequence, the number density of heavy neutrinos fails to follow
the equilibrium concentration reaching at present
\begin{equation}
n = {6\cdot 10^3\over \sqrt{g_\ast}} \Bigl({m_p\over M_{PL}}\Bigr)~ \Bigl({m_p\over
m}\Bigr)~ \Bigl({\rho_c\over 10^{-29} h^2 gcm^{-3}}\Bigr) \cdot \Bigl[\int^{x_f}_0
dxm^2_p (\sigma v)\Bigr]^{-1} cm^{-3}~,
\end{equation}
where $\rho_c = 1.879\cdot 10^{-29} h^2  g~ cm^{-3}$ is the critical density of the
Universe; h is the normalized Hubble constant; M$_{PL} = 1.221 \cdot 10^{19}$ GeV is
the Planck mass; m$_p$ is the proton mass; x = T/m; x$_f$ = T$_f$/m; $\sigma$ is the
annihilation cross section; v is the relative velocity of the neutrino-antineutrino
pair in its center-of-mass frame; g$^\ast$ is the effective number of relativistic
degrees of freedom at T = T$_f$ (bosons contribute 1 to g$^\ast$ and fermions 7/8). The
freeze-out temperature T$_f$ can be computed iteratively from
\begin{equation}
x^{-1}_f = \ln {0.0955 M_{PL} \sigma v m \sqrt{x_f}\over \sqrt{g^\ast}}
\end{equation}
and generally T$_f~\approx$ m/30.

In general case the following processes could lead to the annihilation of heavy
neutrinos in the Universe: $\nu\bar{\nu} \to f\bar{f}$, $W^+W^-$, ZZ, ZH, HH,
however the dominant processes are $\nu\bar{\nu} \to f\bar{f}$ below the
threshold for $W^+W^-$ production and $\nu\bar{\nu} \to W^+ W^-$ above the threshold
[2].

Fig. 1 shows the dependence of cosmological density $\rho_\nu = 2mn$ of heavy
neutrinos as a function of neutrino mass. In the region $m \sim M_Z/2$, the neutrino
density is extremely small as a result of the huge value of the annihilation cross
section at the Z boson pole. When the neutrino mass increases the cross section for
neutrino annihilation drops and this leads to an increase of the neutrino density,
which reaches its maximum value at m $\approx$ 100 GeV. At m $>$ M$_W$ the
annihilation channel into $W^+W^-$ opens and gradually becomes the dominant one, and
since its cross section grows like m$^2$ [2,3] the present neutrino density drops
again.

As it is seen from Fig.1 neutrino density is small in comparison with critical
density. However in the Galaxy neutrino density can be increased by some orders of
magnitude due to neutrino condensation. It was shown [3,4] that at the stage of
galaxy formation the motion of heavy neutrinos in the nonstatic gravitational field
of ordinary matter, which contracts as a result of energy dissipation via radiation,
provides an effective mechanism of energy dissipation for neutrinos. As a
consequence, the contracting ordinary matter induces the collapse of the neutrino
gas and leads to the following significant increase in the neutrino density in the
central part of the Galaxy:
\begin{equation}
n_{0G} \approx n \Bigl({\rho_{0G}\over \rho_U}\Bigr)^{3/4},
\end{equation}
where $\rho_{0G} \approx 10^{-20} g~ cm^{-3}$ is the central density of the matter in
the Galaxy and $\rho_{U} \approx 4~10^{-31} g~ cm^{-3}$ is the density of matter in the
Universe (here we take an upper limit [1] on the density of matter in order to evaluate
neutrino density from below). It is often assumed that the density of dark matter halo
in the Galaxy decreases with the distance from the center according to the law
\begin{equation}
\rho (r) = {\rho_0\over 1+(r/a)^2}~,
\end{equation}
where a (2 kpc $<$ a $<$ 20 kpc) is the core radius of the halo. Assuming that neutrino
density also follow the distribution (4), taking into the minimal value for the
core radius a = 2 kpc and substituting into Eq.(4) the distance $r_{sun} \approx$ 8.5 kpc
from the Sun to the center of the Galaxy we find that the density in the solar
neighborhood is reduced by the factor $\approx$ 19 in comparison with the central
neutrino density (3). Therefore taking into account (3), (4) we obtain that the density
of heavy neutrinos in the solar neighborhood has to satisfy the condition
\begin{equation}
n_{sum} \geq 3.3~10^6 n~.
\end{equation}

Let us apply the results obtained above to the experimental data on the direct search
for WIMPs by WIMP-nucleus elastic scattering. Neutrinos scatter from nuclei by Z boson
exchange and therefore the axial coupling, which only produces small spin-dependent
effects, can be neglected. In the nonrelativistic limit the neutrino cross section due
to coherent scattering on nuclei is given by [5]
\begin{equation}
\sigma_{elastic} \approx {m^2 M^2\over 2\pi (m+M)^2} G^2_F \bar{Y}^2 \bar{N}^2~,
\end{equation}
where $\bar{N} = N - (1 - \sin^2 \theta_w)$ Z; N, Z are numbers of neutrons and protons,
respectively; M is the mass of target nucleus, G$_F$ is the Fermi constant, Y is an
average hypercharge ($\bar{Y} \sim 1$). The nuclear degrees of freedom can be taken into
account by a nuclear form factor [6].

Fig.2 shows an exclusion plot in a cross section of coherent neutrino elastic
interaction with Ge obtained in underground experiments [7] versus heavy neutrino mass.
The theoretical cross section corresponding to heavy neutrino elastic interaction taking
into account the loss of coherence at high masses also is shown in Fig.2. However this
exclusion plot was obtained under the assumption that the heavy neutrinos constitute all
of dark matter with the density $\rho = 0.3$ GeV cm$^{-3}$, which is
5.3$\cdot$10$^{-25}$ g cm$^{-3}$. In order to substitute $\rho$ by $\rho_{Sun}$ and
modify the exclusion plot to the ``real'' case we have to divide the values
corresponding to the bound of the exclusion plot by the ratio
\begin{equation}
\xi = \rho_{Sun}/\rho~.
\end{equation}
The new bound corresponding to the ``real'' neutrino density in the Galaxy is shown also
in Fig.2 by dashed line. As we can see, in this case the existence of very heavy
neutrinos is forbidden in the mass region
\begin{equation}
60  {\rm GeV} \geq m \geq 290  {\rm GeV}~.
\end{equation}
We note that (8) represents the minimal region excluded by Ge underground experiments
because this result has been obtained by using for the astrophysical and cosmological
parameters a conservative set of values. Changing in parameters chosen above leads only
to the extension of the excluded region (8).

Previously less restrictive limits on mass
of heavy neutrinos were obtained [3] from the analysis of the spectrum of electrons in
cosmic rays. However that consideration contained uncertainties related in particular
with poor knowledge of life time of cosmic rays in the Galaxy. Meanwhile bound (8) is
based practically on the only assumption (but well physically motivated) concerning the
condensation of heavy neutrinos in the Galaxy.

Let us note that if Higgs meson exists the bound (8) does not exclude the possibility for
heavy neutrinos to have mass in the region $\vert M_H - m\vert \leq \Gamma_H$, where
$\Gamma_H$ is the width of Higgs meson, because in this case the s-channel
annihilation $\nu\bar{\nu} \to H \to \ldots$ could reduce significantly the neutrino
density in the Galaxy [3].

Recently preliminary results on underground WIMPs search using the annual modulation
signature with large mass NaI(Tl) detectors were published [8]. The overall analysis
has shown that there is an indication on the single crystal response (however as
it was mentioned by the authors [8] only very large exposure would possibly allow to
reach a firm conclusion).

In Fig.3 we have shown qualitatively the region corresponding to the possible signal
[8]  taking into account the ``real'' neutrino density in the Galaxy according to
Eq.(7). It was noted in [8] that the region of the signal is well embedded [6] in the
minimal supersymmetric standard model estimates for neutralino. However as we see from
Fig.3 this case can corresponds also to the elastic scattering by nucleus of NaI of relic
neutrinos with mass between 45 GeV and 50 GeV. This region is consistent with the
present laboratory bound m $>$ 45 GeV.

The confirmation of this event can be obtained in experiments with cosmic rays, by AMS
spectrometer, which is in preparation for the operation on Alfa station. As it was
mentioned in [3,9,10] the detection of an anomalous output of monochromatic positrons
with energy above 45 GeV would be a clear signature of the annihilation of Dirac
neutrinos in the galactic halo because the direct annihilation of Majorana fermions into
electron-positron pair in the Galaxy is severely suppressed. On the other hand the
detection of the irregularity in the continuum spectrum of positrons could be an
indication on the annihilation of neutralinos.

Our approach is easily testable because we assumed only the simplest extended Standard
Model with 4$^{th}$ generation without any ad hoc and fine tuning parameters.

We also note that the search for heavy neutrinos at accelerators [11] in the reaction
$e^+e^- \to \nu\bar{\nu}\gamma$ could give a possibility of analyzing the mass region m
$\sim M_Z/2$, which is difficult for an astrophysical investigation because of small
value of neutrino density but can be important in relation with an event possibly
observed in DAMA experiment [8]. If heavy neutrinos exist there could be also an
interesting hadronless signature for Higgs meson bremsstrahlung production at
accelerators $e^+e^- \to ZH \to Z\nu\bar{\nu} \to 1^+ 1^- \nu\bar{\nu}$ and this mode
could be the dominant one.

In conclusion we emphasize that it seems that only complex investigations including
underground experiments, accelerator searches, and astrophysical investigations can
clarify the physical nature of dark matter.

This work was supported by part by the scientific and educational center ``Cosmion''.
M.Yu.K. and R.V.K. are grateful to I Rome University for hospitality and support. We
thank D.Prosperi for very interesting discussions.
\vfill\eject

\noindent
{\bf FIGURE CAPTIONS}

\begin{itemize}
\item{Fig.1} Cosmological density of heavy neutrinos as a function of neutrino mass.

\item{Fig.2} Mass exclusion plot for interaction of dark matter particles with Ge. Dashed
line represents the boundary [7] corresponding to the case when neutrinos constitute all
dark matter ($\rho =0.3$ GeV cm$^{-3}$). Solid line corresponds to ``real'' neutrino
density in the Galaxy. The curve near the bottom of the plot shows the predicted cross
section for a Dirac neutrino.

\item{Fig.3} Region corresponds qualitatively to the region of possible dark
matter signal [8] at m $>$ 45 GeV. Line intercepting the region shows the predicted cross
section for a Dirac neutrino.
\end{itemize}
\vfill\eject

%

\end{document}